\documentclass[prd,aps,floats,twocolumn,nofootinbib,superscriptaddress]{revtex4}
\usepackage{commath}
\pdfoutput=1
\usepackage[usenames, dvipsnames]{color}
\usepackage{graphicx, array}
\usepackage{multirow}
\usepackage{graphics}
\usepackage{graphics,epsfig}
\usepackage{amssymb}
\usepackage{amsmath}
\usepackage{ulem}
\usepackage{multirow}
\usepackage{subfigure}
\usepackage{bm}
\usepackage{txfonts}
\usepackage{cancel}
\usepackage{url}
\usepackage{hyperref}
\usepackage{lipsum}
\usepackage{graphicx}

\newcolumntype{C}[1]{>{\centering\arraybackslash}m{#1}}

\renewcommand{\vec}[1]{\boldsymbol{\mathbf{#1}}}

\begin{document}
\input{epsf}

\title{Pairwise kSZ signal extraction efficacy and optical depth estimation}

\author{Yulin Gong}
\author{Rachel Bean}
\affiliation{Department of Astronomy, Cornell University, Ithaca, NY 14853, USA}
\author{Patricio A. Gallardo}
\affiliation{Kavli Institute for Cosmological Physics, University of Chicago, 5640 S. Ellis Ave., Chicago, IL 60637, USA}
\author{Eve M. Vavagiakis}
\affiliation{Department of Physics, Cornell University, Ithaca, NY 14853, USA}
\author{Nicholas Battaglia}
\affiliation{Department of Astronomy, Cornell University, Ithaca, NY 14853, USA}
\author{Michael Niemack}
\affiliation{Department of Physics, Cornell University, Ithaca, NY 14853, USA}
\affiliation{Department of Astronomy, Cornell University, Ithaca, NY 14853, USA}

\begin{abstract}
We determine the efficacy of the kinematic Sunyaev-Zel'dovich signal extraction pipeline, using pairwise kSZ measurements, in recovering unbiased estimates of the signal and inference of the associated optical depth. We consider the impact of cluster co-alignments along the line of sight, the modeling of baryonic clustering, and the presence of diffuse gas, as well as instrument beam convolution and noise. We demonstrate that two complementary approaches, aperture photometry, and a  matched filter, can be used to recover an unbiased estimate of the cluster kSZ signal and the associated optical depth. Aperture photometry requires a correction factor accounting for the subtraction of signal in the annulus while the matched filter requires a tuning of the signal template profile. We show that both of these can be calibrated from simulated survey data. The optical depth estimates are also consistent with those inferred from stacked thermal SZ measurements. We apply the approaches to the publicly available Atacama Cosmology Telescope (ACT) data. The techniques developed here provide a promising method to leverage upcoming kSZ measurements, from ACT, Simons Observatory, CCAT, and CMB-S4 with spectroscopic galaxy surveys from DESI, Euclid, and Roman, to constrain cosmological properties of the dark energy, gravity, and neutrino masses.
\end{abstract}
\maketitle
\section{Introduction}
The cosmic microwave background (CMB) radiation has provided exquisite constraints on the standard cosmological model \citep{Planck:2013oqw, Planck:2015mrs, Planck:2018vyg, WMAP:2003ivt, WMAP:2010qai, WMAP:2012nax} through the correlations in the primary CMB anisotropies, those imprinted in the CMB at its inception during recombination. Secondary anisotropies, induced by the interaction and scattering of the CMB photons after the last-scattering surface, as they travel through cosmic structures, also now present an invaluable complementary probe to discern the composition of matter, properties of gravity, and the history of cosmic evolution. This includes the polarization signatures imprinted during the epoch of reionization, gravitational lensing of the CMB, the linear integrated Sachs-Wolfe effect (ISW), the non-linear Rees-Sciama evolution (RS), and the temperature modifications induced by Sunyaev-Zel'dovich (SZ) effects as the CMB passes through the gas in large-scale structure(LSS) of the Universe.

These late-time imprints on the CMB left by the LSS of the Universe provide insights into the evolution of cosmic structure and the properties of dark energy, the unknown quantity responsible for the accelerated expansion of the universe \citep{Peebles:2002gy}, and of dark matter and neutrinos. Each of these is only detectable indirectly through inferring their impacts on the clustering of baryonic matter, the motions of cosmological objects, and gravitational lensing \citep{Mueller:2012kb, Mueller:2014dba, Mueller:2014nsa}.

The SZ effect can be principally separated into two effects: thermal and kinematic \cite{Sunyaev:1970eu, Sunyaev:1972eq, Sunyaev:1980nv} (and see reviews \cite{Birkinshaw:1998qp, Carlstrom:2002na}). The thermal Sunyaev-Zel'dovich (tSZ) effect is a spectral distortion of the CMB photons through inverse Compton scattering off electrons in hot gas. The kinematic Sunyaev-Zel'dovich (kSZ) effect is a consequence of a galaxy cluster’s line of sight motion relative to the Hubble flow. When CMB photons traverse through a galaxy cluster, the peculiar motion of the cluster (with respect to the rest frame of the CMB) creates a Doppler shift in the CMB. The tSZ effect has a characteristic frequency dependence that provides a natural approach to detection using multi-frequency observations. By comparison, the kSZ effect is an order of magnitude smaller than the tSZ effect when looking at clusters and does not have a similarly distinctive frequency dependence. As a result, its detection presents a greater challenge. 

In this paper, we focus on a common two-point statistic used to extract the kSZ signal from groups and clusters, the pairwise momentum statistic \citep{Davis:1982gc}. In this, the inherent infall of pairs of clusters towards each other due to gravity leads to a correlation in the resulting kSZ Doppler signatures that can be leveraged to extract the kSZ from the CMB. A number of other techniques have also been proposed to extract the kSZ signal. These include velocity reconstruction stacking \citep{ACTPol:2015teu, Tanimura:2020une, Tanimura:2022fde, DES:2023mug}, kSZ tomography statistics \citep{Smith:2018bpn, Sato-Polito:2020cil}, and projected-field estimators \citep{Dore:2003ex, Hill:2016dta, Ferraro:2016ymw, Kusiak:2021hai, Bolliet:2022pze, Patki:2023fjz}. Other approaches include a velocity matched filter \citep{Li:2014mja}, cross-correlation with 21cm data \citep{Li:2018izh, Sato-Polito:2020cil}, a local measurement for an individual cluster \citep{Sayers:2013ona}, evidence from signatures in the CMB power spectrum \citep{George:2014oba}, and angular redshift fluctuations \cite{Hernandez-Monteagudo:2019epd, Chaves-Montero:2019isa}.

The kinematic SZ effect, with its sensitivity to the cosmic velocity fields, encodes rich information about tracers of the gravitational field, with the potential to provide insights into cosmology \citep{Haehnelt:1995dg, Diaferio:1999ig, Aghanim:2001yu, Bhattacharya:2006ke}. This includes potential constraints on the dark energy and modified gravity models \citep{DeDeo:2005yr, Hernandez-Monteagudo:2005xtx, Bhattacharya:2007sk, Kosowsky:2009nc, Keisler:2012eg, Ma:2013taq, Mueller:2014nsa, Sugiyama:2016rue, Alonso:2016jpy, Howard:2022klw} and constraints on the sum of the neutrino masses \citep{Mueller:2014dba}. Projected field approaches have also been shown to have the potential to constrain non-Gaussianity \citep{Munchmeyer:2018eey, Jolicoeur:2023tcu, Sullivan:2023qjr}. Constraints on optical depth are necessary in order to constrain cosmology from the pairwise kSZ signal \citep{DES:2016umt, Soergel:2016cqh, DeBernardis:2016pdv, Flender:2016cjy, Battaglia:2016xbi, Vavagiakis:2021ilq, Calafut:2021wkx, SPT-3G:2022zrq, Madhavacheril:2019buy}.

The kSZ signal is sensitive to the halo thermodynamics \citep{Battaglia:2017neq, Amodeo:2020mmu} and the modeling of the intra-cluster gas \citep{Ostriker:2005ff, Bode:2009gv, Shaw:2010mn, Flender:2016cjy}. Different modeling of cluster baryonic physics using hydrodynamical simulations and analysis of large-scale simulations have been performed for kSZ statistics \citep{Sehgal:2009xv, Flender:2015btu, Soergel:2017ahb, Stein:2020its, Hadzhiyska:2023cjj} and the effect of central galaxy miscentering in clusters has also been studied \citep{Flender:2015btu, Calafut:2017mzp}. 

CMB data gathered by high-resolution telescopes, the Atacama Cosmology Telescope (ACT)  \citep{Swetz:2010fy}, $Planck$, and the South Pole Telescope \citep{Carlstrom:2009um} (SPT) have enabled measurements of the pairwise kSZ effect.  This includes detections of the pairwise kSZ effect in ACT \citep{Hand:2012ui, DeBernardis:2016pdv} and $Planck$  \citep{Planck:2015ywj} CMB data using spectroscopically selected galaxy catalogs from the Baryon Oscillation Spectroscopic Survey \citep{SDSS:2012gam} (BOSS). A detection has also been made using CMB data from SPT and the $Planck$ with photometrically selected galaxies from the Dark Energy Survey \citep{DES:2016umt}. The evolution of intergalactic gas is also studied using the kSZ effect with BOSS galaxies, and SDSS quasars as tracers \citep{Chaves-Montero:2019isa}.

Upcoming CMB experiments, including the Simons Observatory \citep{SimonsObservatory:2018koc}, CCAT Observatory \citep{Aravena:2019tye}, and CMB-S4 \citep{CMB-S4:2016ple}, in combination with large scale structure survey data from the Vera Rubin Observatory (Large Synoptic Survey Telescope, LSST) \citep{LSSTScience:2009jmu},  Dark Energy Spectroscopic Instrument(DESI) \citep{DESI:2013agm}, Euclid Space Telescope \citep{Amendola:2016saw}, and Roman Space Telescope \citep{Akeson:2019biv, Eifler:2020vvg}, will provide higher precision data over a broader range of redshifts to expand and improve the pairwise kSZ analysis.

A central objective for using kSZ pairwise momentum measurements is to obtain an estimate of the cluster pairwise velocity correlation from which we can make inferences about the underlying cosmological model. This requires both an accurate measurement of the pairwise kSZ momentum and an accurate way to estimate the optical depth of the clusters and groups used for the pairwise momentum measurements. 

One tangible approach to measure the pairwise velocity correlation is to combine measurements of the optical depth from tSZ measurements with the pairwise kSZ momentum. In order to confidently implement this, we need to verify that the tSZ and kSZ measurements can be combined in this way without introducing biases.

In two recent works \cite{Vavagiakis:2021ilq, Calafut:2021wkx}, a comparison of the optical depths estimated from the stacked tSZ and pairwise kSZ measurements for the same sample of cluster and groups sources was undertaken. For the kSZ data, the mass-averaged optical depth was inferred using aperture photometry from the pairwise kSZ momentum by comparing with an analytical cluster pairwise velocity correlation prediction based on a Planck best-fit cosmology and linear theory. The optical depth derived from pairwise kSZ measurements in this way was found to be smaller than that obtained from the stacked aperture photometry tSZ maps.

Motivated by the clear cosmological value of measuring the pairwise velocity correlation, and the apparent optical depth discrepancies found in the previous work, the work in this paper centers on understanding and validating our ability to accurately measure the pairwise kSZ momentum using not only the aperture photometry approach but also develop an alternative complementary approach using matched filtering. Explicitly, we focus on the optical depth estimates obtained from the kSZ aperture photometry and matched filter pipelines and how they are related to the underlying physical line of sight cluster optical depth. These analyses will then allow us to determine how the optical depth estimates from other approaches, such as from the tSZ, can be used to extract out the pairwise velocity correlation.

In this work, the simulated datasets afford us full knowledge of the halo properties such as the halo mass and halo peculiar velocities, with related kSZ and tSZ temperatures. These enable us to precisely calculate the pairwise kSZ momentum and pairwise velocity correlations as well as the stacked tSZ temperatures. We use this knowledge to study how the pairwise momentum estimates obtained from the kSZ signal extraction methods compare to the true simulated pairwise kSZ momentum, and related to the velocity correlation under different assumptions. We also, in tandem, gain an understanding of the optical depth inferred from these kSZ based methods for a set of cluster and group samples that mirror those we are planning to observe in the context where we have full knowledge of the cluster velocities. We can also ascertain how the optical depth estimates compare with the tSZ signal extracted from the same samples. In this way, we explore the most effective ways to use kSZ pairwise momentum observations and tSZ measurements to make inferences about the large-scale structure peculiar velocities and related cosmological constraints.

The rest of the paper is organized as follows: In section \ref{sec:formalism}, we describe the simulation maps and catalogs that we use in this work and introduce the formalisms for the matched filter, aperture photometry, pairwise estimator, covariance estimation, and mass-averaged optical depth inference. In section \ref{sec:analysis}, we present our findings, analyzing the effects on kSZ signal extraction of overlapping structures and diffuse gas along the line of sight, the results of the recovered signal with the different extraction methods, and the application of simulation validated approaches to ACT DR5 and SDSS data. In section \ref{sec:conc}, we conclude with a discussion of the findings and implications for future research.

\section{Formalism }
\label{sec:formalism}

The key objective of this work is to calibrate the relationship between kSZ signals extracted from the CMB and the underlying cluster properties. With this, we can determine the potential to use SZ measurements to make cosmological inferences. In section \ref{sec:SZ_signal} we outline how the SZ signals are related to the cluster properties. The simulations used in this work are described in \ref{sec:data}. The filtering techniques and statistics used for SZ signal extraction are described in section \ref{sec:filter} and \ref{sec:Phat} respectively. The formalism for connecting the SZ signals to cosmological inferences is described in \ref{sec:tau}.

\subsection{SZ signals}
\label{sec:SZ_signal}
The kSZ signal is induced by the Doppler shift in the CMB  by the motion of a galaxy cluster along the line of sight. To first order, it is frequency independent and is given by:
\begin{equation}
\label{eq:kSZ}
    \frac{\delta T_{kSZ}}{T_0} = - \int_{los} \frac{v_{los}}{c}\, n_e\,\sigma_{T}\,dl,
\end{equation}
where $T_0$ = 2.726K is the average CMB temperature, $v_{los}$ is the line of sight peculiar velocity, c is the speed of light, $n_e$ is the electron number density, and $\sigma_T$ is the Thomson cross-section. A positive line of sight peculiar velocity, $v_{los}$, corresponds to a cluster moving away from the observer, so induces a negative kSZ effect. 

The tSZ signal is frequency dependent, given by:
\begin{equation}
\label{eq:tSZ}
    \frac{\delta T_{tSZ}(v)}{T_0} = f_{v} \, y,
\end{equation}
where $y$ is the Compton-y parameter, and in the non-relativistic limit, $f_\nu$ = x coth(x/2) - 4 where x = $h\nu$/($k_B T_{CMB}$) \citep{Itoh:1999wt}. The Compton-y parameter can be defined as:
\begin{equation}
\label{eq:comptony}
    y = \frac{\sigma_T}{m_e c^2} \int_{los} P_e dl,
\end{equation}
where $m_e$ is the electron mass and $P_e$ is the radial electron pressure. 

\subsection{Datasets}
\label{sec:data}
The simulations used for the analysis are described in \ref{sec:sims}. We discuss how the simulated primary CMB and instrument beam and noise are modeled in \ref{sec:noise}.  Finally, we also analyze CMB data from the Atacama Cosmology Telescope (ACT) described in \ref{sec:ACT}. 
 
\subsubsection{SZ simulations}
\label{sec:sims}

We utilize data products of simulated kinematic Sunyaev-Zel'dovich maps from \citep{Flender:2015btu}\footnote{\url{https://www.hep.anl.gov/cosmology/ksz.html}}(herein the Flender simulations), to determine the efficacy of the SZ signal extraction pipeline, and to understand the connection between the observed signal and the inferred optical depth.  

The Flender SZ maps were generated via post-processing from an N-body simulation output \citep{Habib:2014uxa} that adopts a WMAP7 cosmology \citep{WMAP:2010qai} with $3200^3$ particles simulated in a (2.1 Gpc)$^3$ volume and a mass resolution of roughly $10^{10} M_{\odot}$. The halos are identified using a friends-of-friends (FoF) algorithm \citep{Knebe:2011rx}. 

For each N-body simulated halo, the entire particle distribution is assigned a single (mass-weighted average) peculiar velocity that is provided in the halo catalog. This halo velocity is used to calculate the pairwise velocity correlation for the analysis in this work.

The kSZ signal predicted from three models of the intra-cluster gas clustering is simulated from the particle level data: Model 1 (herein FL1) assumes baryons trace the dark matter and includes the diffuse kSZ component contributed from filaments and the inter-galactic medium (using the positions and velocities of particles outside haloes inferred from the simulation output). Model 2 (FL2) uses a gas prescription from the Shaw model \citep{Shaw:2010mn} in which the gas initially is modeled as a Navarro-Frenk-White (NFW) profile \citep{Bartelmann:1996hq} and then follows the hydrostatic equilibrium model. Model 3 (FL3) follows Model 2 using the Shaw model and also includes kSZ from diffuse gas. A full-sky tSZ map is also simulated using the Shaw gas prescription.

We consider two mass-selected cluster samples: a high-mass sample (herein FLhi) with a mass range of $1 \times 10^{14} M_{\odot} < M_{200} < 2.5 \times 10^{14} M_{\odot}$ and a lower mass sample (herein FLlo) with mass range $1 \times 10^{13} M_{\odot} < M_{200} < 1 \times 10^{14} M_{\odot}$. Both samples are selected in the redshift range $0.35<z<0.65$. The FLlo sample is intended to be reflective of the luminosity-cut ``L61" sample used in the recent ACT DR5 - SDSS DR15 kSZ and tSZ analyses \citep{Vavagiakis:2021ilq, Calafut:2021wkx}. The L61 sample is selected using luminous red galaxies (LRGs) as tracers of the galaxy groups and clusters. An empirical relationship between the tracer galaxy luminosity and the host group/cluster halo mass is used to determine the halo mass \citep{Chabrier:2003ki}. The FLlo sample has $\langle z\rangle= 0.52 \pm 0.09$, $\langle M\rangle= (2.38 \pm 1.65) \times 10^{13} M_{\odot}$, and the L61 sample has $\langle z\rangle = 0.51 \pm 0.13$, and an inferred mass range of $\langle M\rangle = (2.46 \pm 1.55) \times 10^{13} M_{\odot}$. The FLlo sample is selected to have a comparable number of halos to L61, FLlo contains 205,651 halos and L61 comprises 213,070 target galaxies (used to identify the clusters)\citep{Vavagiakis:2021ilq, Calafut:2021wkx}. We consider a FLhi sample containing 150,530 halos which is a much bigger sample size than the equivalent one from the ACT DR5 data. Our intent with the high mass sample is not to do a comparison with ACT but instead to study the differences in signal extraction for high and low mass cluster properties. 

The Flender simulated maps have a resolution of 0.43$^\prime$, providing a comparable resolution to the ACT data, and kSZ models that include both well-motivated baryonic clustering and diffuse gas. Through selecting the FLlo sample to have a comparable mean redshift, mean cluster mass, and sample size to the ACT DR5 sample, the intent is for it to provide an accurate characterization with which to test our signal extraction process and relate it to ACT analyses. 
\subsubsection{Simulated CMB maps} 
\label{sec:noise}
 
 To model an expected observed foreground-cleaned CMB signal, we augment the kSZ maps with simulated primary CMB anisotropies and instrument noise and convolve with the instrument beam. Random primary CMB map realizations are generated from a CMB angular power spectrum obtained from CAMB \citep{Lewis:1999bs} with the appropriate WMAP cosmological model parameters. Instrument noise is simulated per pixel as a Gaussian distribution with a white noise level of 10 $\mu$K-arcmin, corresponding to the noise coming from ACTPol \citep{Swetz:2010fy} and forecasted for Simons Observatory at 145~GHz \citep{SimonsObservatory:2018koc}. 

The effects of instrument beams are incorporated by considering two beam profiles: the ACT DR5 beam (herein the ACT Beam) \citep{Naess:2020wgi} \footnote{\url{https://lambda.gsfc.nasa.gov/product/act/actpol_prod_table.html}} and, for generality, a Gaussian beam of FWHM=1.4$^\prime$ in accordance with the expected beam scale for Simons Observatory at 145~GHz \citep{SimonsObservatory:2018koc}.

\subsubsection{ACT CMB Maps}
\label{sec:ACT}
In addition to the simulated maps described above, we also consider the analysis and interpretation of the kSZ and tSZ signals extracted from the co-added ACT DR5 150~GHz temperature map \citep{Naess:2020wgi}. This combines ACT observations at a single frequency with the $Planck$ PR2 data release at 143~GHz \citep{Planck:2015mrs}. This map is referred to as the DR5 f150 map and covers about 21,100 square degrees of the sky, and has a resolution of 0.5$^\prime$.

\subsection{Signal extraction}
\label{sec:filter}

We consider two signal extraction approaches: aperture photometry, the most commonly employed technique, as outlined in \ref{sec:ap} and a new calibrated application of matched filtering for kSZ signal extraction, in \ref{sec:mf}. 

\subsubsection{Aperture Photometry}
\label{sec:ap}
Aperture photometry is a measurement of flux magnitude obtained by calculating the average temperature within a given disk and then subtracting the average temperature of the sky background contribution from an annulus just outside the disk of equal area. The aperture photometry temperature obtained for a disk of angular radius $\theta_{AP}$ in the direction $\hat{n}$, can thus be expressed as
\begin{equation}
    T_{AP}(\hat{n},\theta_{AP}) =  T_{disk}(\hat{n},\theta_{AP}) -  T_{ring}(\hat{n},\theta_{AP}).  
\end{equation}

We use the software pipeline built in \citep{Calafut:2021wkx}\footnote{\url{https://github.com/patogallardo/iskay}}. Postage stamps of 45' are excised around the locations of each target halo in the catalog. The postage stamps are repixelized to 10 times finer resolution before ${T}_{disk}$ and ${T}_{ring}$ are calculated to allow the circular disk and ring boundaries to be well-characterized and avoid coarse pixelization effects leading to bleeding of contributions between the disk and ring and the ring and the area exterior to it.

We use aperture photometry for both kSZ and tSZ signal extraction. For the tSZ, this can be connected to the inferred AP-derived Compton-y parameter using (\ref{eq:tSZ}).
\subsubsection{Matched Filter}
\label{sec:mf}
A Matched Filter (MF) is a method to detect the presence of a template signal from an unknown background signal. It is also a linear filter process that maximizes the signal-to-noise ratio (SNR) for a known template signal embedded in unknown background noise. Given we have prior knowledge of the template profile and shape of the kSZ signal, we can use the matched filter method to extract the kSZ signal from clusters and CMB.

For a given template profile,  $\tau$, that is embedded in a noisy background with power spectrum, $P(k)$, a Fourier space matched filter, $\Psi$, can be constructed as  \citep{Haehnelt:1995dg, Melin:2006qq}:
\begin{equation}\label{eq:mf}
    \Psi(k) = \sigma^2 \frac{\tau(k)B(k)}{P(k)},
\end{equation}
where $B(k)$ is the beam function of the CMB experiment, and $\sigma^2$ is the variance of the filter which is given by:  
\begin{equation}\label{eq:sigma}
    \sigma^2 = \left[\int \left(\frac{\tau(k)B(k)}{2\pi}\right)^2 \frac{d^2k}{P(k)}\right]^{-1}.
\end{equation}

We approximate the power spectrum of the total noise as  $P(k)=P_{CMB}(k)B(k)^2 + P_{noise}(k)$, where we approximate the noise as the power spectrum of a combination of the primary CMB anisotropies, $P_{CMB}$, with the other noise contributions $P_{noise}$, which we model here as instrument noise. 

For our analysis, we use the projected Navarro-Frenk-White (NFW) profile \citep{Bartelmann:1996hq} as our template for profile $\tau$, given by:
\begin{equation}\label{eq:nfw}
    \tau(x) = \frac{A}{x^2-1} 
    \begin{cases}
    1 - \frac{2}{\sqrt{1-x^2}}\,\text{tanh}^{-1}\sqrt{\frac{1-x}{x+1}}& \text{0 $<$ $x$ $<$ 1} \\
    0 & \text{$x$ = 1} \\
    1 - \frac{2}{\sqrt{x^2-1}}\,\text{tan}^{-1}\sqrt{\frac{x-1}{x+1}} & \text{$x$ $>$ 1},
    \end{cases}
\end{equation}
where A is the normalization factor of the profile, $x = c_{200}\theta/\theta_{200}$, where $c_{200}$ and $\theta_{200}$ are the concentration and angle of the cluster measured at $R_{200}$ respectively. Alternatively, $x$ can also be parameterized with $x = \theta/\theta_{s}$, where $\theta_{s} = \theta_{200}/c_{200}$ is the scale angle of the NFW profile. $\theta_{s}$ typically can range from 0 to 20$^\prime$, and for instance, $\theta_{s}$ can equal to 0.9$^\prime$ for a cluster of mass $10^{14} M_{\odot}$.

The matched filter is applied by convolving the filter in Fourier space with postage stamps cut out at the location of each cluster. The temperature, $T_{MF}(\hat{n},\theta_s)$, is then measured as the average temperature of the pixels within 2.1$^\prime$ from the filtered stamp centered in direction $\hat{n}$.

\subsection{kSZ pairwise momentum estimator}
\label{sec:Phat}

The temperature, $\delta T_i$, is used to define the kSZ temperature around the target galaxy, with direction $\hat{n}_i$ and redshift $z_i$, used as a proxy locator of the center of the $i^{th}$ cluster, 
\begin{equation}
    \delta T_i(\hat{n}_i,z_i) = T(\hat{n}_i) - \overline{T}(\hat{n}_i,z_i,\sigma_z),
\end{equation}
where $T(\hat{n}_i)$ is the kSZ signal, that can be $T_{disk}$, $T_{AP}$, or $T_{MF}$.

A redshift-smoothed temperature, $\overline T$, is subtracted to remove potential sources of redshift-dependent noise that might mirror a pairwise kSZ signal,
\begin{equation}
    \overline{T}(\hat{n}_i,z_i,\sigma_z)=\frac{\sum_j T(\hat{n}_i)w(z_i,z_j,\sigma_z)}{\sum_jw(z_i,z_j,\sigma_z)},
\end{equation}
where
\begin{equation}
    w(z_i,z_j,\sigma_z) = \exp\left(-\frac{(z_i-z_j)^2}{2\sigma_z^2}\right),
\end{equation}
and we use $\sigma_z$ = 0.01 in accordance with \citep{Li:2017uin}.

The observed pairwise momentum estimator can be written in terms of CMB temperatures,
\begin{equation}\label{eq:pob}
    \hat{p}(r,z)= -\frac{\sum_{ij}(\delta T_i - \delta T_j)c_{ij}}{\sum_{ij}c_{ij}^2}.
\end{equation}
This differences the temperatures in pairs of clusters, $\delta T_i$ and $\delta T_j$, across all pairs in the catalog sample. The differences for all pairs in the same comoving radial separation bin, centered at $r=|\vec{r}_{ij}| = |\vec{r}_i - \vec{r}_j|$,  are summed, and weighted by a factor  $c_{ij}$ relating to the geometry of the cluster pair orientation relative to the line of sight, given by
\begin{equation}
    c_{ij}=\hat{r}_{ij} \, \dot{}\,\frac{\hat{r_i}-\hat{r_j}}{2}=\frac{(r_i-r_j)(1+cos \, \theta)}{2\sqrt{r_i^2\,+r_j^2\,-2 r_i r_j cos \, \theta}}.
\end{equation}
with $\theta=\hat{r}_i.\hat{r}_j$, the angle between the two position unit vectors $\hat{r}_i$ and $\hat{r}_j$. 

To facilitate comparison with  \citep{Vavagiakis:2021ilq, Calafut:2021wkx}, we consider bins in radial separation, $r$,  of equal width of 10 Mpc from 0 to 150 Mpc, and four unevenly spaced bins centered on 175, 225, 282.5 and 355 Mpc.

\subsection{Optical depth estimation}
\label{sec:tau}
Our principal aim is to measure the pairwise velocity, $V(r)$, from the measurements of pairwise momentum which can be modeled as:
\begin{equation} \label{eq:pairwise}
    \hat{p}_{model}(r,{\tau}) = -\frac{T_{CMB}}{c}\,V(r)\,{\tau},
\end{equation}
where ${\tau}$ is an effective mass-averaged optical depth over the cluster sample and $V(r)$ is the pairwise peculiar velocity, modeled in a parallel way to the temperature pairwise momentum to maintain consistent weightings in (\ref{eq:pob}):
\begin{equation}\label{eq:Vhat}
    V (r)= -\frac{\sum_{ij}(v_i - v_j)c_{ij}}{\sum_{ij}c_{ij}^2},
\end{equation}
with individual cluster velocity $v_i$. Note that in this work we directly use the halo velocity provided by the simulation catalog. However, in real observation, the pairwise velocity estimator, $V(r)$, is obtained by using the theoretical prediction from linear theory. 

A likelihood of ${\tau}$ is then determined by $\chi^2$, for example for the pairwise statistic, 
\begin{equation}
    \chi^2({\tau}) = \sum_{ij} \delta \hat{p}_i({\tau})\,{C}_{ij}^{-1}\,\delta\hat{p}_j({\tau}),
\end{equation}
where $i$ here denotes the bin in $r$ (not an individual galaxy cluster), and 
\begin{equation}
    \delta\hat{p}_i({\tau}) = \hat{p}_{model}(r_i,{\tau})-\hat{p}_{obs}(r_i).
\end{equation}
Here $\hat{p}_{obs}$ and $\hat{p}_{model}$ represent the measured and modeled pairwise momentum estimate respectively, and $C^{-1}_{ij}$ is the inverse of the covariance matrix obtained through bootstrap analysis. 

We use the bootstrap re-sampling strategy to estimate the covariance for $\hat{p}$. The temperature decrements, $\delta T$, of galaxy positions are randomly reassigned, with repeated values allowed. The covariance matrix, $C_{ij}$, is estimated from the variance of the pairwise momentum estimators calculated for each of 1,000 repeated random reassignments.

The best fit $ \tau$ is obtained by finding the minimum $\chi^2$. We also calculate the probability of obtaining a higher $\chi^2$ value, the Probability-To-Exceed (PTE),
\begin{equation}
    PTE=\int_{\chi_{min}^2}^{\infty} \chi^2(x)dx. 
\end{equation}

For the simulations, as a consistency check, we also obtain an average optical depth estimate from fitting the individual halo temperature decrements, \{$T_{kSZ}$\}, to the halo LOS peculiar velocities \{$v_{los}$\}. From (\ref{eq:kSZ}),
\begin{eqnarray} \label{eq:linear_fit}
    \frac{\delta T_{kSZ,i}}{T_0} = -\frac{v_{los,i}}{c}\tau_i.
\end{eqnarray}

Note that these analyses are inferring a cylindrical estimation of the optical depth which is akin to the real observation that the measurement includes contributions from intervening structures along the line-of-sight (see, for example, \citep{Soergel:2017ahb, Hadzhiyska:2023cjj}).

\section{Analysis and Results}
\label{sec:analysis}

In section \ref{sec:effect_overlapping} we consider the implications of cluster two-halo term, line of sight co-alignment, and diffuse gas contributions for the kSZ pairwise signal and derived optical depth. In \ref{sec:ap_result} we look at the connection between kSZ pairwise momentum and peculiar velocities for the aperture photometry extraction method. A scaling relation for the optical depth with the aperture photometry derived tSZ signal is also discussed. A 2.1$^\prime$ disk and aperture size is used to facilitate comparison with \citep{Calafut:2021wkx, Vavagiakis:2021ilq}. In section \ref{sec:mf_result} we present a new proposed approach for kSZ signal extraction using a calibrated matched filter. In section \ref{sec:ACT_results} we apply the approaches to the ACT DR5 data.
 
\subsection{Effects of two-halo term and overlapping line-of-sight clusters and diffuse gas}
\label{sec:effect_overlapping}

\begin{figure}[t!]
\includegraphics[width = \columnwidth]{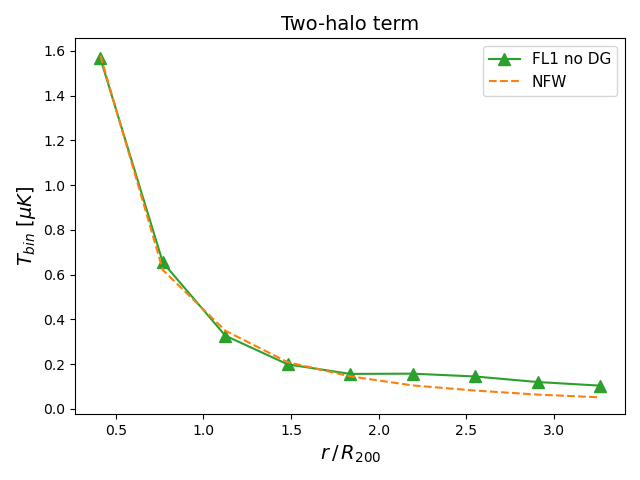}
\caption{The average radial temperature profiles for a diffuse gas free FL1 map (FL1 no DG) and a sample of halos each having a proximate halo within 50 Mpc [green,triangle] and analytical model of signal with an NFW baryonic profile [orange, dashed line]. The sample has an average $M_{200}$ of $2.3 \times 10^{13} M_{\odot}$ and average redshift of 0.51.} 
\label{fig:two_halo_term}
\end{figure}

In a simplistic modeling of the kSZ signal predicted for clusters, we might consider the kSZ signal from an isolated, spherically symmetric cluster with no further contributions to the kSZ signal along the line of sight. In reality, we have to consider the impact of three modifications to this simple model. First, contributions from second halos within the correlation scale of linear velocity field, roughly 50 Mpc  \citep{Amodeo:2020mmu},  which can contribute
to the measured signal of the primary halo, the two-halo term, that has been carefully modeled \citep{Amodeo:2020mmu, Hill:2017tua, Vikram:2016dpo}. Second, there might be more than one cluster or group coincidentally aligned along the line-of-sight (LOS) whose kSZ anisotropies will also contribute to the observed signal. Third, we expect there to be additional diffuse gas contributions in the inter-cluster medium that could also induce kSZ anisotropies in the CMB that add to those from the clusters. In this section, we consider the potential effects of these two additional contributions to the measured pairwise signal.

In this section, we consider the $T_{disk}$ temperature of the kSZ signal, to assess the impact of two-halo term, and line-of-sight correlated IGM and cluster co-alignment, and then also discuss the impact when aperture photometry and matched filter are used. 

To investigate the kSZ signal contribution from proximate halos, the two-halo term, we use the FL1 kSZ model map, which includes both a kSZ signal, based on the baryons tracing the dark matter with an NFW cluster density profile, and diffuse inter-cluster gas, and create a diffuse gas-free version of FL1 (FL1 no DG) by subtracting an estimate of the diffuse gas map obtained by differencing the FL2 and FL3 maps. We randomly select 10,000 halos from the catalog that each have a proximate halo within 50 Mpc. This sample has an average $M_{200}$ of $2.3 \times 10^{13} M_{\odot}$ and average redshift of 0.51. We calculate the average angular temperature profiles for the clusters measured from the FL1 kSZ model with no diffuse gas but including the two-halo contribution, and compare it to the analytical model of the signal where the electron number density $n_e$ is modeled by the NFW profile for the single primary halos. The difference between these is the two-halo contribution. We show the comparison in Fig.~\ref{fig:two_halo_term} and find that the contribution of the two-halo term only significantly contributes to the signal on scales above $2R_{200}$, consistent with the results in \citep{Amodeo:2020mmu}. In our work we use a 2.1$^\prime$ aperture size (with the AP outer radius of 2.96$^\prime$) which is smaller than the angular scale subtended by 2$R_{200}$, $\sim 3.2^\prime$ for a cluster of $3\times10^{13} M_{\odot}$ at z = 0.5,  thus we expect the potential correlation of the two-halo term will not bias our measured signal in this analysis.

\begin{figure}[t!]
\includegraphics[width = \linewidth]{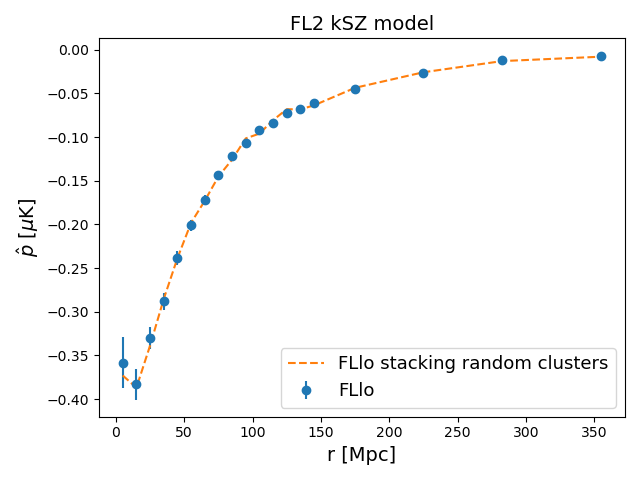}
\caption{Pairwise kSZ momentum estimator [blue points] for the FLlo halo sample and FL2 kSZ model is compared to the signal for the same dataset with the maps having had the equivalent of two additional halos randomly co-aligned along the line of sight [orange dashed line].} 
\label{fig:effect_coalign_fl2}
\end{figure}

We assess the potential impact of co-aligned clusters along the LOS using the FLlo sample. For each primary halo, we search for all other halos in the simulated datasets that are co-aligned with it within the 2.1$^\prime$ radius disk region. We find that there are 1.7 additional halos, on average, overlapping with the primary halo within 2.1$^\prime$ disk along the line of sight direction and the distribution of the overlapping objects is consistent with a Poisson distribution. 

To assess how the addition of signals from co-alignment impacts the pairwise momentum statistic, we mimic co-alignment by stacking random halo objects on top of our primary halos. We first randomly select halos from the full catalog (beyond FLlo) and create postage stamp cutouts for them from the FL2 kSZ map (which just includes the intra-cluster gas signal and no diffuse gas). We then randomly stack these FL2 kSZ cutouts on top of the FL2 kSZ map at the location of the primary halo sources in the FLlo sample with centers aligned. On average, each of the FLlo halos has an additional two cutouts stacked on it.

Figure~\ref{fig:effect_coalign_fl2} compares the pairwise signal calculated for both the original FL2 map and the map with the additional cutouts added. We find that the addition of random LOS co-alignments does not create any systematic bias in the resulting kSZ pairwise momentum signal. The estimator is indeed unbiased in the presence of uncorrelated LSS as initially designed.

\begin{figure}[t!]
\includegraphics[width = \linewidth]{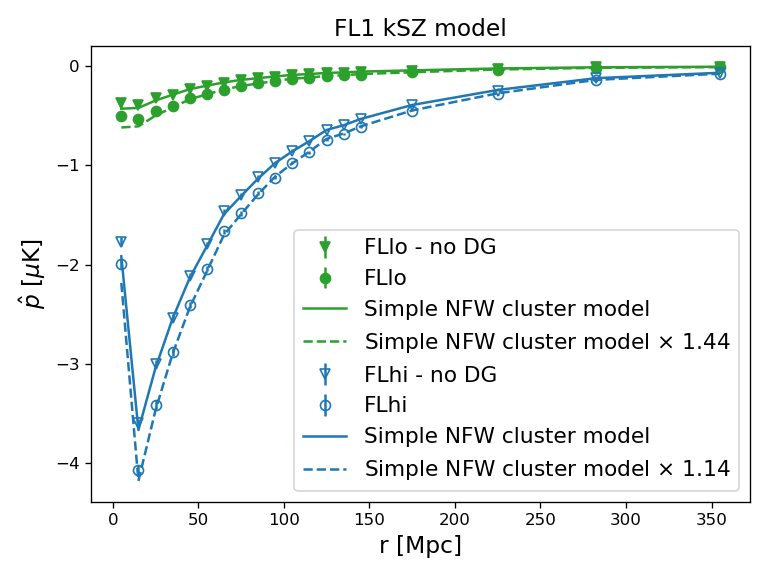}
\caption{Pairwise kSZ momentum estimators using $T_{disk}$ as the input
for the low mass, FLlo, [filled] and high mass, FLhi, [hollow] halo samples for the FL1 kSZ map [circle points] and a diffuse gas free (no DG) FL1 map  [triangle points]. The theoretical kSZ prediction for isolated spherical clusters with an NFW baryonic profile [full line] is consistent with the signal from the diffuse gas free map. The signal when the diffuse gas is included is systematically boosted [dashed line] relative to the NFW theory prediction by  44\% and 14\%  for the low and high mass samples respectively.}
\label{fig:ksz_AP_effect}
\end{figure}

\begin{figure*}
\includegraphics[width = 0.49\linewidth]{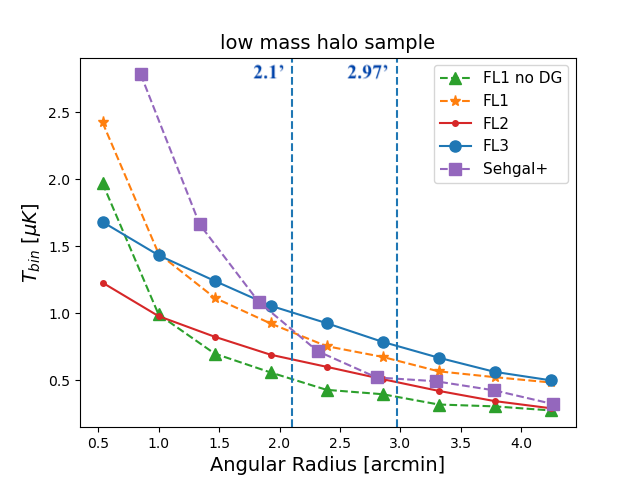}
\includegraphics[width = 0.49\linewidth]{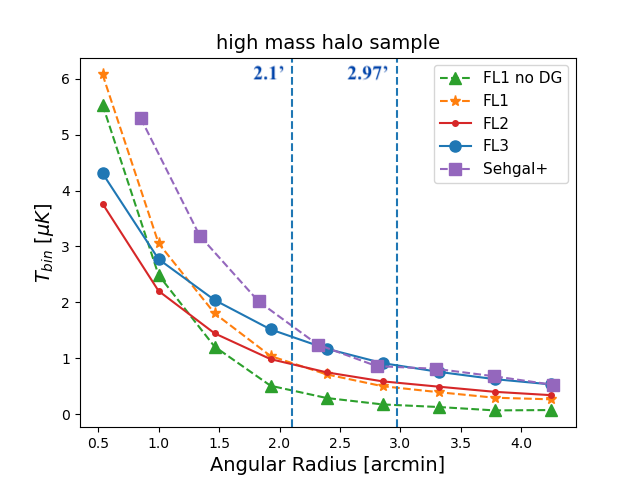}
\caption{The average radial temperature profiles for halos from low-mass [left] and high-mass [right] halo samples for the FL1 kSZ model with no diffuse gas (“FL1- no DG”) [green,triangle] and using FL1 [orange,star], FL2 (no diffuse gas) [red,small circle] and FL3 [blue,big circle] kSZ signal
models, and Sehgal+ kSZ model[purple,square]. The inner disk and ring annulus regions (respectively within 2.1$^\prime$ and $\sqrt{2}  \times $ 2.1$^\prime$=2.97$^\prime$) are also indicated. The profile is binned by roughly 0.5 arcminutes, e.g. the value at 0.5 arcminutes represents the average between 0 and 0.5$^\prime$.} 
\label{fig:radial_profile}
\end{figure*}

To determine the impact of diffuse inter-cluster gas on the pairwise signal, we consider the FL1 kSZ model map and its diffuse gas-free version of FL1. We also create a theoretical cluster-only kSZ map by estimating the signal for isolated clusters at each halo location in the FLlo sample using (\ref{eq:kSZ}), where the electron number density $n_e$ is modeled by the NFW profile.

In Fig.~\ref{fig:ksz_AP_effect}, we compare the predicted pairwise signals for these two maps. While the diffuse gas free map is consistent with the simple theoretical prediction from assuming a spherical NFW-profile cluster, the diffuse gas provides a significant pairwise signal amplification to the kSZ signals coming from the clusters alone. This is equivalent to a scale-independent signal boost of 1.44 $\pm$ 0.02 and 1.14 $\pm$ 0.01 for the FLlo and FLhi samples respectively on the diffuse-gas-free signal. We interpret this effect to be a result of the diffuse gas in the locale of the clusters having peculiar velocities that become correlated with the gravitationally dominant cluster. As a result, the diffuse gas produces a coherent additional signal to the cluster gas. This demonstrates that the diffuse gas contributes a significant amount to the effective optical depth in the kSZ signal and must be accounted for; it is insufficient to model the optical depth purely assuming isolated spherical clusters. 

When we consider the pairwise momentum from aperture photometry filtering we find the signal boosts are 1.37 $\pm$ 0.02 and 1.03 $\pm$ 0.01 for the FLlo and FLhi samples respectively, while for matched filtering, different scale angles are used to account for the change of signal template profile due to the diffuse gas. This reflects that diffuse gas needs to be considered when using the two filtering approaches. 

\subsection{Aperture Photometry}
\label{sec:ap_result}
\subsubsection{Impact of AP subtraction on kSZ signal measurement}
\label{sec:ap_result_1}

In using the aperture photometry approach, a common perception might be that the aperture would be chosen to wholly enclose the target object so that annulus subtraction is simply removing the extraneous background or foreground signals unrelated to the object. In reality, the aperture size may be smaller than the target object radius for a number of reasons. The signal associated with the target cluster decreases as we move outwards from the center, so choosing a smaller aperture size that focuses on the brightest inner region maximizes the signal-to-noise measurement. A fixed aperture (rather than one based on a potentially inaccurate mass/size estimate) makes analysis and error estimation simpler, however, we are considering objects of inherently different angular size (due to variations in physical size and redshift) so a fixed aperture choice will not perfectly match all halos simultaneously. Given the choice of aperture size, we must factor in that the annulus subtraction will not only remove extraneous signals but also some of the target cluster kSZ signals as well \cite{Soergel:2017ahb}. 

To be able to relate the AP pairwise signal to other optical depth estimates, from theoretical or simulation predictions, or tSZ measurements, for example, we need to know how much signal the AP approach removes so we can correct for that. In this section, we determine these correction factors and how they depend on the cluster sample masses and the way in which the baryon density profile is modeled. We calibrate these corrections using the pure simulated kSZ alone, and then in the subsequent section determine how these can be applied to more realistic simulations including primary CMB and instrument noise.

We first assess the effect of the aperture photometry subtraction by directly considering the angular temperature profiles for the clusters. Figure~\ref{fig:radial_profile} shows the average radial temperature profiles for the FLlo and FLhi halo samples for kSZ signals predicted for the FL1, FL2, and FL3 models.  The annulus region for each of the three models, $2.1^\prime<\theta<2.97^\prime$, still contains a significant fraction of the cluster signal. In the FL1 model, the density profile falls off more steeply relative to the other models, as a result, the annulus subtracts off a smaller fraction, 30\% and 28\% of the disk signal for the FLlo and FLhi samples, respectively. In FL2 the baryon distribution is relaxed relative to the CDM density profile, through assuming that the gas, initially modeled as an NFW profile, the gas in the model rapidly re-distributes itself into hydrostatic equilibrium. This softens the density and temperature profile and leads to the annulus subtraction having a larger impact relative to FL1, removing 51\% and 39\% of the disk signal for the low and high mass samples. In FL3, the inclusion of diffuse gas increases the temperature profile across the scales included in both the disk and the annulus, particularly for the low-mass sample. This leads to the annulus subtraction removing 53\% (FLlo) and 49\% (FLhi) of the disk signal. 

It's clear that the specific modeling of the gas prescription impacts the overall kSZ simulation using a hydrostatic equilibrium model.  We also analyze the kSZ simulations built on those described in Sehgal et al. \cite{Sehgal:2009xv} and updated by Colin Hill \citep{SimonsObservatory:2018koc} (herein Sehgal+ simulation). These simulations use a gas prescription  \cite{Ostriker:2005ff, Bode:2009gv} that, like FL3, builds on a hydrostatic equilibrium model and includes, for example, the effects of star-formation rate, non-thermal pressure support and feedback from AGNs. The main difference is the inclusion of radially dependent non-thermal pressure support in FL3 which is not included in the Sehgal+ simulations. As discussed in \cite{Shaw:2010mn}, in which a comparison of the two models is presented, the inclusion of this component does lead to notable differences in the baryon distribution in the halos. In particular, for a low-mass and high-mass halo sample in the Sehgal+ simulation comparable to FLlo and FLhi, we find that the radial temperature profiles are more peaked in the inner region of the halo than that for FL3 in Fig.~ \ref{fig:radial_profile}. This leads to the overall fraction of signal removed from the AP annual subtraction being lower, removing 24\% and 27 \% of the disk signal for the low-mass and high-mass samples respectively.

\begin{figure}[t!]
\includegraphics[width = \linewidth]{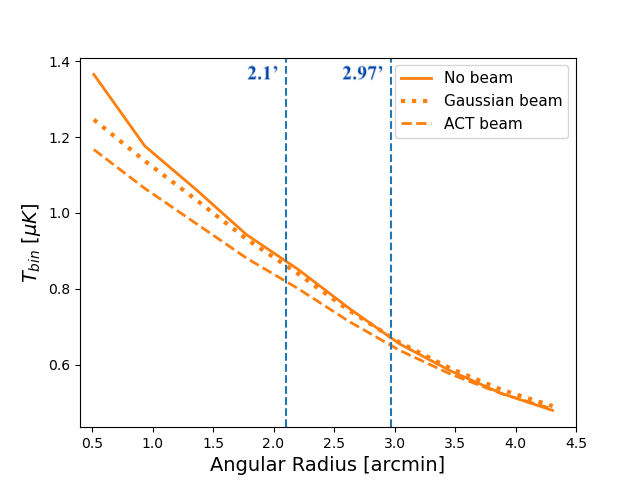}
\caption{The average radial temperature profile for the FLlo halo sample and the FL3 kSZ model under convolution with no beam [full], a Gaussian beam [dotted], and the ACT beam [dashed]. The profile is binned by roughly 0.5 arcminutes, e.g. the value at 0.5 arcminutes represents the average between 0 and 0.5$^\prime$. }
\label{fig:radial_beam}
\end{figure}

Just as the baryonic physics modeling has an impact on the expected radial profile, so too does the instrument beam, which smooths out the temperature fluctuations through its characteristic geometry. In Fig.~\ref{fig:radial_beam}, we show the impact of the beam smoothing on the average radial profile, for the FLlo halo sample and FL3 kSZ model, for both a Gaussian beam and the ACT beam. In each case, the smoothing decreases the amplitude in the disk and enhances it in the annulus, with the ACT beam smoothing the map slightly more than the Gaussian one. This smoothing leads to the annulus subtraction being more pronounced for the beam-convolved maps. Here the annulus subtraction removes 64\% for the ACT beam in comparison to 53\% when no beam smoothing is applied.

\begin{figure}[t!]
\includegraphics[scale=0.55]{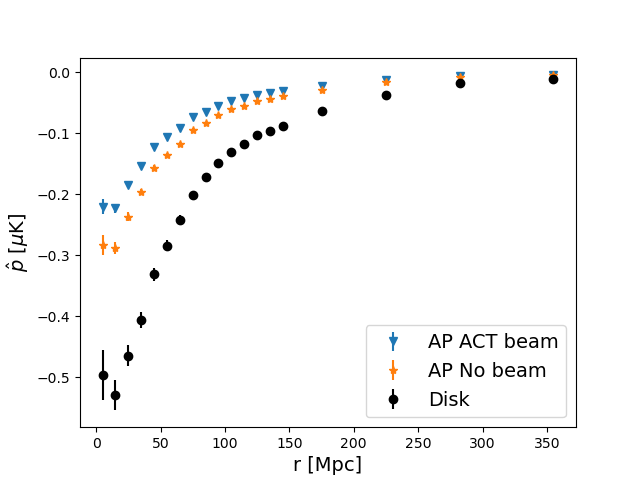}
\caption{Pairwise kSZ momentum estimators for the FLlo halo sample and  FL3 kSZ model, with no CMB or noise added. The estimator is calculated using the $T_{disk}$ (with no aperture photometry) [circle] and from $T_{AP}$ with no beam [star] and the ACT beam [triangle]. The $1 \sigma$ variance (of the kSZ signal alone) is also included for each scenario.
}
\label{fig:pureksz_model3_ap_pairwise}
\end{figure}

While the radial temperature profile provides an intuitive way to consider the way in which the AP subtraction affects the signal amplitude, we seek to measure the impact of AP subtraction on the pairwise momentum statistics directly. As already discussed, the pairwise statistic is effective at removing extraneous signals that are uncorrelated with the velocity correlation of each pair of target clusters and is, therefore, a cleaner signal to analyze. In Fig.~\ref{fig:pureksz_model3_ap_pairwise} we show  $\hat{p}_{AP}$ for the FLlo sample of the FL3 kSZ model and beam scenarios. By calculating the pairwise velocity statistic for the FLlo and FLhi samples, we can obtain estimates of the inferred effective optical depth from both a pairwise measurement using (\ref{eq:pairwise}) for the disk temperature directly, $\hat{p}_{disk}$  and from $AP$ measurements, $\hat{p}_{AP}$.

\begin{table*}[]
\begin{tabular}{|C{5em}|C{6.em}|C{6.em}|C{6em}|C{6.em}|C{6.em}|C{6em}|}
\cline{2-7}
\multicolumn{1}{c|}{}& \multicolumn{3}{c|}{FL3 map + FLlo halo sample} & \multicolumn{3}{c|}{FL3 map + FLhi halo sample}           
\\ \hline
Beam &  $10^4{\tau}_{AP}$  & $10^4{\tau}_{disk}$  &  $A_{\tau}$ & $10^4{\tau}_{AP}$ & $10^4{\tau}_{disk}$  &  $A_{\tau}$ 
\\ \hline
None & 0.88 $\pm$ 0.01 & 1.89 $\pm$ 0.02  &  2.15 $\pm$ 0.02  & 3.12 $\pm$ 0.02 & 6.12 $\pm$ 0.03  & 1.96 $\pm$ 0.01
\\ \hline
 Gaussian  & 0.78 $\pm$ 0.01 & 1.81 $\pm$ 0.02  &  2.42 $\pm$ 0.03 & 2.80 $\pm$ 0.02 & 5.87 $\pm$ 0.03  & 2.19 $\pm$ 0.01
\\ \hline
 ACT & 0.69 $\pm$ 0.01 & 1.66 $\pm$ 0.02  & 2.74 $\pm$ 0.03  & 2.48 $\pm$ 0.01 & 5.40 $\pm$ 0.03  & 2.47 $\pm$ 0.02
 \\ \hline
\multicolumn{7}{c}{}
\end{tabular}
\caption{The inferred mass-averaged optical depth values derived from the pairwise momentum signals for the disk temperature, ${\tau}_{disk}$,  and that from aperture photometry, ${\tau}_{AP}$, as inputs,  using a 2.1'  disk/aperture size, and their ratio, $A_\tau={\tau}_{disk}/{\tau}_{AP}$ for the FL3 kSZ map (without CMB and noise) and different beam assumptions and the low and high mass-selected halo samples, FLlo [left] and FLhi [right], respectively.}
\label{tab:tauratio}
\end{table*}

In Table~\ref{tab:tauratio}, we summarize the inferred effective optical depths from the disk and from the AP measurements for the different halo samples and kSZ models.  We also provide the multiplicative attenuation factors, $A_\tau$ that would be required to shift the mean of a theoretical estimate of $\tau_{disk}$ to compare with the AP-inferred value, $\tau_{AP}$, in each case.

In general, the ACT beam-derived has the largest signal removal with the annulus subtraction, consistent with what we see in Fig.~\ref{fig:radial_beam}, where the ACT beam leads to the greatest smoothing of the signal amplitude across the AP scales. The low-mass sample has a bigger subtraction factor than the high-mass sample. This is as expected given the radial temperature profiles are less peaked for the low-mass sample, and the diffuse components are a  larger fractional component of the total signal, so the annulus contributes a greater fraction of the overall halo signal. This can also be seen from Fig. \ref{fig:radial_profile} that the radial temperature profiles are less peaked for the low-mass sample and the profile drops slowly while it drops sharply for the high-mass sample in the disk region. In the meanwhile, the annulus has a comparable temperature amplitude between the low and high-mass samples leading to a larger fractional subtraction for the low-mass sample. Therefore, low mass halos should yield larger subtraction factors.
\begin{figure}
\includegraphics[width = \columnwidth]{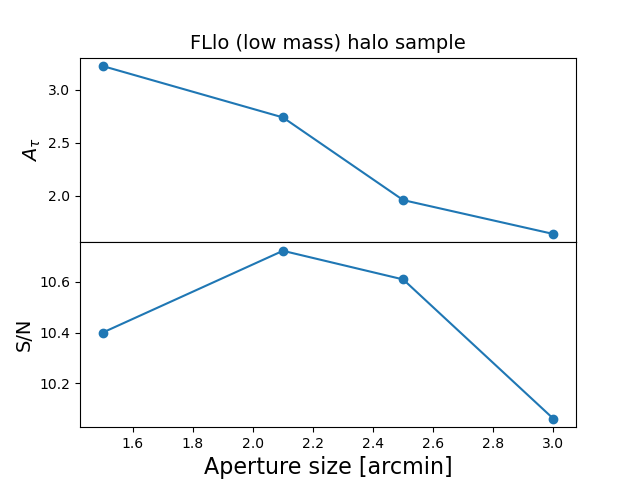}
\caption{The attenuation factor, $A_{\tau}$ [upper], and S/N [lower] as a function of aperture size for the FLlo sample + FL3 map and the ACT beam.}
\label{fig:atte_SNR}
\end{figure}

\begin{figure*}
\includegraphics[width = \linewidth]{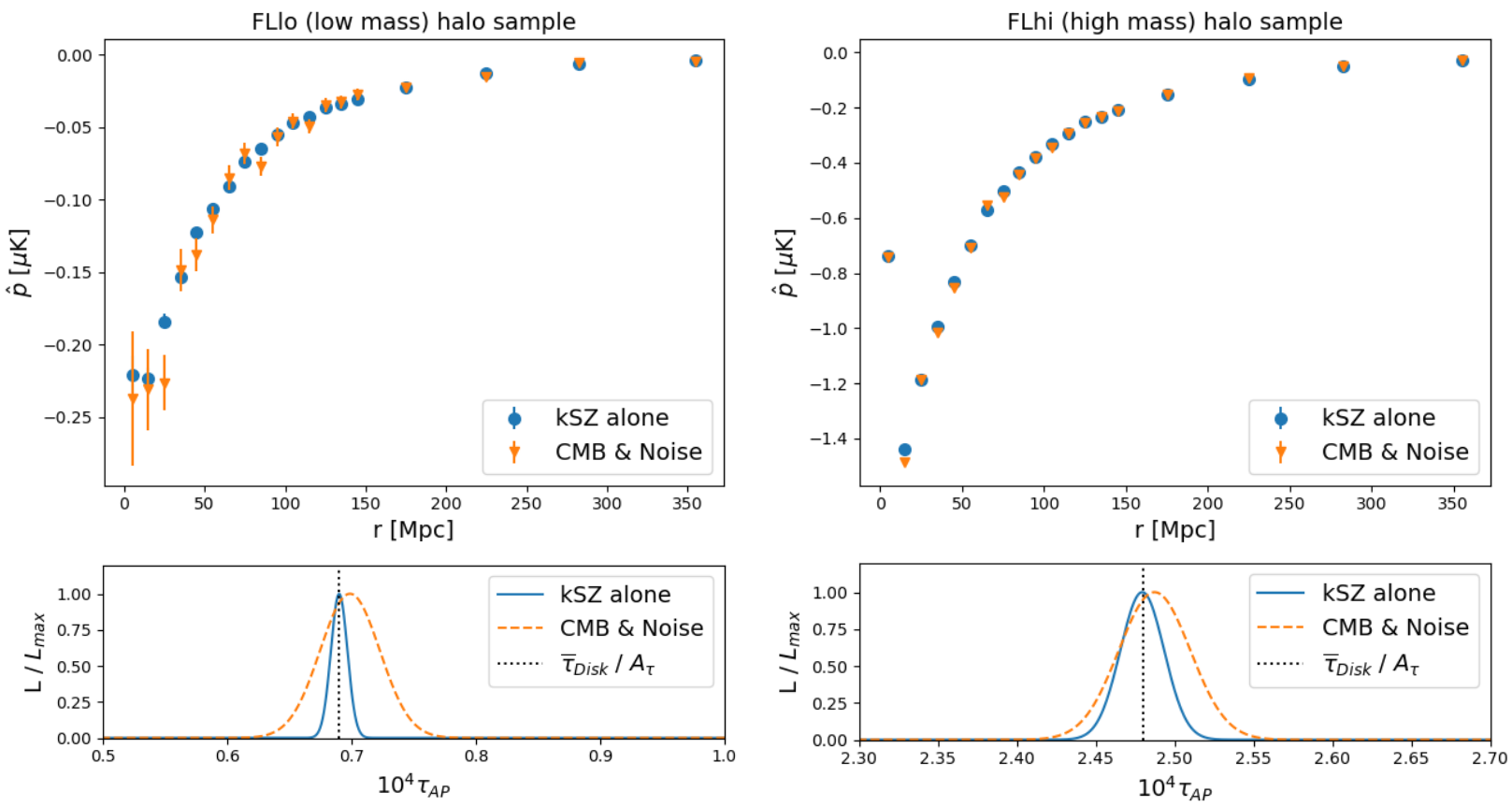}
\caption{The mean and 1-$\sigma$ error on the mean of the aperture photometry (AP) pairwise momentum estimator [upper]  and normalized likelihoods for AP-obtained $\tau_{AP}$ estimates [lower] derived from the FL3 kSZ maps, for the kSZ alone  [blue] and in combination with 10 independent CMB+instrument noise realizations [orange] for the FLlo [left] and FLhi [right] halo samples with the ACT beam. The optical depth for the 2.1$^\prime$ disk is also shown with the attenuation factor, $A_\tau$, applied to enable comparison with the AP measurement.}
\label{fig:actbeam_ap}
\end{figure*}

Larger apertures have less annulus subtraction but lower signal-to-noise (S/N) in the inner disk signal, smaller apertures offer higher S/N in the disk but also higher subtraction levels from the outer ring. To assess if there is an ``optimal" aperture size that maximizes the S/N of the inner disk while minimizing the signal subtracted by the outer ring annulus, we consider three other aperture sizes 1.5$^\prime$, 2.5$^\prime$, and 3$^\prime$ along with the fiducial 2.1$^\prime$. In Figure \ref{fig:atte_SNR}, we summarize the attenuation factor and S/N for each of the aperture sizes. We find that the S/N is peaked around 2.1$^\prime$ (S/N = 10.7) and the S/N decreases as the aperture size increases above 2.6$^\prime$. On the basis of this, we conclude it is not preferable to select a larger aperture size with the intent of avoiding a correction for the aperture photometry attenuation as it would also reduce the S/N. 

In addition to the Flender simulations, we analyzed the low-mass sample in Sehgal+ simulations with an ACT beam convolution, and find the different baryonic modeling leads to a lower $A_{\tau}= 2.41 \pm 0.02$ than the FLlo sample. This is consistent with the Bode model used in these simulations predicting a more peaked radial temperature profile than the Shaw model used in Flender. The Sehgal+ simulation also has a lower resolution (0.87$^\prime$) than the Flender simulation. We find that degrading the FL3 to the 0.87$^\prime$ resolution does change the fraction removed by AP, with the lower resolution leading to more smoothing and slightly increasing $A_\tau$ from 2.74 to 2.88. This change is smaller than that induced by the baryonic modeling differences but still indicates the importance of selecting a comparable resolution to the maps being analyzed in determining the impact of the AP annulus subtraction.

In summary, the optical depth derived from aperture photometry is attenuated due to the annulus subtraction of the kSZ signal. The disk optical depth estimates can be connected to the AP-derived optical depth estimates via a factor that depends on aperture size, beam profile, and the gas model. 

\subsubsection{kSZ signal recovery from Aperture Photometry in the presence of CMB and instrument noise}
\label{sec:AP_CMB_Noise}

In this section, we consider how well AP can extract out an unbiased estimate of the kSZ signal, and the optical depth, once instrument noise and the primordial CMB are included, which in combination are orders of magnitude greater than the kSZ signal itself.

We create 10 realizations of the CMB and instrument noise modeled from the $Planck$ CMB spectrum \citep{Planck:2018vyg} and the ACT noise spectrum \citep{Henderson:2015nzj} and add them to the FL3 kSZ map and convolve with the ACT beam. We filter out large-scale signals above 30 arcminutes (i.e. $\ell < $ 360) and keep, the kSZ relevant, small-scale signals below 15 arcminutes (i.e. $\ell > $ 720), following \citep{Tanimura:2022fde}. We then calculate the mean pairwise signal across the realizations and estimate the error by bootstrapping one realization.

In Fig.~\ref{fig:actbeam_ap}, we show the pairwise momentum correlations for both high and low-mass halo samples for the maps with CMB and noise included and compare it with the signal for the pure kSZ map. 

The estimator is indeed, again, unbiased in the presence of primary CMB and instrument noise, while, as is expected, the statistical variance is increased. This is also reflected in the likelihoods of the inferred optical depths, shown in the lower panels of Fig.~\ref{fig:actbeam_ap}. The mean optical depth from the kSZ+CMB+noise simulations is unbiased relative to that inferred from the pure kSZ. We find that for the FLlo sample with ACT beam $10^4\overline{\tau}_{AP}=0.69\pm0.01 $ for the kSZ alone and $10^4\overline{\tau}_{AP}=0.70\pm 0.02$ when primary CMB and instrument noise is added, with uncertainties two times larger. For the FLhi sample, the results are $10^4\overline{\tau}_{AP}=2.48\pm 0.02$ for the kSZ alone and $10^4\overline{\tau}_{AP}=2.49\pm0.03$ when primary CMB and instrument noise are added, with uncertainties 1.5 times larger. This comparatively smaller decrease in the uncertainty for the larger mass sample is consistent with the noise being a smaller fraction of the mean signal amplitude. 

As a cross-check available in simulations, we can also obtain an average optical depth estimate from fitting the individual halo temperature decrements to the halo LOS peculiar velocities using (\ref{eq:linear_fit}). We find $10^4\overline{\tau}_{AP}=0.68\pm 0.03$  and $10^4\overline{\tau}_{AP}=2.50\pm 0.04$ for FLlo and FLhi respectively when primary CMB and instrument noise is added. This method gives a comparable estimation of the average optical depth compared with the $\hat{p}$ fit with slightly greater errors consistent with the pairwise statistic removing some of the CMB and diffuse signal that are uncorrelated with the target halos.

\subsubsection{tSZ signal extraction and scaling relations for the optical depth}
\label{sec:tsz_result}

\begin{figure}
\includegraphics[width = \columnwidth]{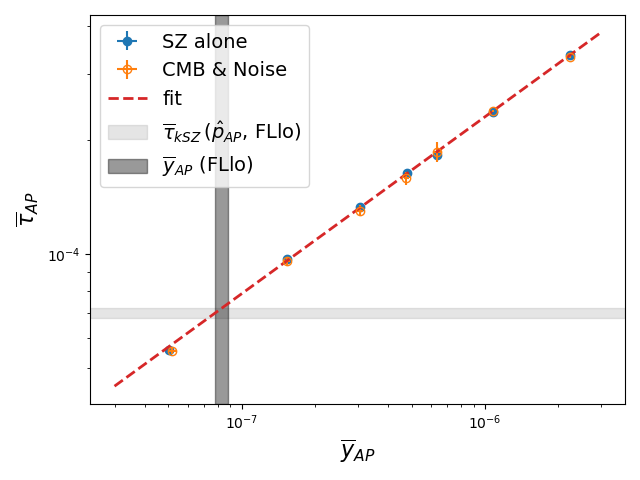}
\caption{The $\overline{y}_{AP}-\overline{\tau}_{AP}$ scaling relation derived from AP measurements of the tSZ and FL3 kSZ maps are shown with the kSZ \& tSZ alone [blue] and from 10 independent CMB+instrument noise realizations [orange] for the mass bins across the combined FLlo and FLhi samples. The dashed line shows the best scaling relation fit [red] with SZ alone, for which $ln \, \overline{\tau}_{AP} = -9.45 + 0.47 \ ln(\overline{y}_{AP}/10^{-7})$. The mean and 1-$\sigma$ ranges for the AP-derived $\tau$ estimates obtained from the pairwise kSZ signal ($10^4 \overline{\tau}_{AP} = 0.70 \pm 0.02$) [light gray] and the average Compton-y ($\overline{y} = 0.83 \pm 0.05$) [dark gray] for the FLlo sample, with primary CMB and noise added, are also shown.} 
\label{fig:y_tau}
\end{figure}

We also consider how the tSZ signal extracted using AP connects to the optical depth estimates inferred from the kSZ signal. Following \citep{Vavagiakis:2021ilq}, we measure the tSZ signal by first stacking the tSZ cutout stamps of each halo and then applying the aperture photometry on the stacked stamps to measure the average $\overline{y}$ signal. The tSZ signal, like the kSZ, is attenuated by the AP annulus subtraction. We directly find a $\overline{y}_{AP}-\overline\tau_{AP}$ scaling relation without explicitly finding a multiplicative attenuation factor to connect the tSZ signal when aperture photometry is employed to that of the disk. We first divide the FLlo+FLhi sample into logarithmically spaced mass bins. For each mass bin, we fit a linear relationship between the individual halo kSZ temperature decrements with AP to the halo LOS peculiar velocities, and determine the AP-derived optical depth from the slope of this linear fit using (\ref{eq:linear_fit}). We then measure the tSZ signal using AP on the stacked tSZ cutout stamps of each mass bin. In Figure \ref{fig:y_tau}, we present the binned $\overline{y}$ measurements and $\overline{\tau}$ estimates we obtained as described above. Following \citep{Battaglia:2016xbi, Soergel:2017ahb}, we model the $\overline{y}_{AP}-\overline\tau_{AP}$ relation, from the SZ maps alone without CMB and noise, using,
\begin{equation}\label{eq:ytau}
\ln \overline{\tau}=\ln\overline{\tau}_0+\alpha \ln\left(\frac{\overline{y}}{\overline{y}_0}\right),
\end{equation}
for which we find, with $y_0=10^{-7}$, $\alpha =0.47 \pm 0.02$ and $\ln\overline\tau_0=-9.45 \pm 0.01$.

Note the fit parameters are dependent on the mass ranges, and redshifts employed and whether $T_{AP}$ or $T_{disk}$ is fitted and the chosen aperture size. For comparison, Soergel et al.\citep{Soergel:2017ahb} reported $y_0=10^{-6}$, $\alpha =0.40 \pm 0.01$ and $\ln\overline\tau_0=-7.94 \pm 0.01$, using AP and the virial radius of each object as the aperture size, for a sample $M_{500} > 3 \times 10^{13}h^{-1} M_{\odot}$ and $0.38 < z < 0.57$. Battaglia et al.\citep{Battaglia:2016xbi} also provides a scaling relation, using a 1.8$^\prime$ disk size without AP applied, for a sample $M_{500} > 10^{14} M_{\odot}$ at z = 0.5, for which $y_0=10^{-6}$, $\alpha =0.49 \pm 0.04$ and $\ln\overline\tau_0=-6.40 \pm 0.09$. 

When we include the 10 independent realizations of the CMB and instrument noise, similar to the kSZ signal, we find that the AP subtraction is effective at isolating the tSZ signal. We find that for the FLlo sample $\overline{y}_{AP} = 0.82 \pm 0.01$ for the tSZ alone and $\overline{y}_{AP} = 0.83 \pm 0.05$ when primary CMB and instrument noise are included. Using the scaling relation, this implies a tSZ inferred optical depth of  $10^4\overline{\tau}_{AP}^{tSZ} = 0.72 \pm 0.02$ when primary CMB and instrument noise are included for the FLlo sample. This is in good agreement with optical depth inferred from the pairwise kSZ signal for the same sample, $10^4\overline\tau_{AP}=0.70\pm0.02$. In summary, using the tSZ signal with AP and an appropriately calibrated scaling relation can provide an alternative and accurate measurement of the optical depth estimates in the presence of primary CMB and detector noise.

\subsection{Matched Filter}
\label{sec:mf_result}
As a complementary approach to aperture photometry, we employ a second signal extraction method, matched filtering, to measure the kSZ signal. Similar to aperture photometry, we convolve the kSZ maps with an assumed beam before applying the matched filter. 

We focus our attention here on using the MF approach to recover the signal from the FL3 kSZ model, which is the most realistic baryonic model, though we also found the approach can equally be applied to the other models.

In many works, the matched filter is used to recover the peak kSZ temperature at the center $x_0$ that 
\begin{equation}\label{eq:peak}
    T_{x_0} \approx \int \Psi^t(x_0 - x') M(x') dx',
\end{equation}
where $T_{x_0}$ is the kSZ temperature at $x_0$, $\Psi$ is the filter described in (\ref{eq:mf}) and is normalized to unity at the center, superscript t indicates a transpose, and M is the input signal. In this work, however, we focus on using the matched filter to recover the average kSZ amplitude within a disk of a given angular radius. In other words, our work is similar to minimizing the mean squared error,
\begin{equation} \label{eq:mse}
    \text{MSE} = \frac{1}{n}\sum_{x=0'}^{x=2.1'} [T_x - \int \Psi^t(x-x') M(x') dx']^2,
\end{equation}
with a different normalization of the filter $\Psi$ such that the filtered halo radial profile is as close as possible to the true halo radial profile within 2.1$^\prime$.
\begin{figure}[t!]
\includegraphics[width = \linewidth]{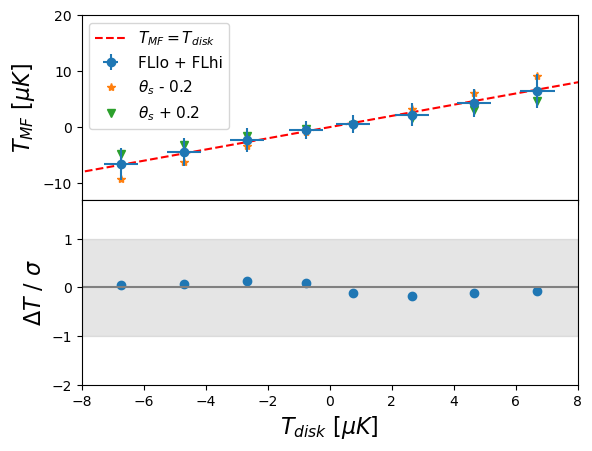}
\caption{[Upper] A comparison of the average disk temperature within 2.1$^\prime$ measured from the kSZ-only map, $T_{disk}$, and that recovered from the beam-convolved kSZ-only map with matched filter, $T_{MF}$, using the selected $\theta_s$ for each of FLlo and FLhi samples,  with 1$\sigma$ error [blue circle]. MF results using, $\theta_s\pm 0.2$ [orange star and green triangle] are also shown. [Lower] The difference between the mean predicted and expected temperatures ($\Delta T = T_{MF} - T_{disk}$ ) scaled relative to the error in the mean.}
\label{fig:temp_comp}
\end{figure} 
As discussed in \ref{sec:mf}, we use the projected NFW profile as our template signal, parameterized with the scale angle $\theta_s$. We find that we can effectively extract out the signal by solely using the single parameter $\theta_s$, which can both adjust the width and the vertical scale of the filter. We do not need to employ two parameters: the normalization, $A$, and $\theta_s$, and therefore set $A=1$. We calibrate the scale angle by randomly picking 10,000 objects from the halo sample and tuning the scale angle such that the average temperature within 2.1$^\prime$ recovered with the matched filter is a best-fit estimate of the actual average disk temperature within a 2.1$^\prime$ aperture. We find that projected NFW profile scale radii of $\theta_s$ = 0.97$^\prime$ and 0.91$^\prime$ work well for the low and high mass (FLlo and FLhi) mass samples, respectively. 
When combining the FLlo and Flhi samples, we separately apply MF to each sample, using their respective $\theta_s$, and then analyze the $T_{MF}$ samples in combination.

In Fig.~\ref{fig:temp_comp}, we compare the $ T_{disk}(2.1^\prime)$ measured from a kSZ-only map with the $ T_{MF}(2.1^\prime)$ that is recovered from the Gaussian beam-convolved kSZ-only map with matched filter. We find a good direct proportionality relationship between $T_{MF}$ and $T_{disk}$ with MF providing an unbiased estimate of $T_{disk}$ within the statistical sensitivity of our analysis, with any errors or biases between the two temperatures averaged to zero when they are propagated to the pairwise statistics. Given the MF filtered signal has a Fourier profile proportional to $(\tau_k B_k)^2$, one might be concerned that it would introduce a bias,  as shown in Fig.~\ref{fig:temp_comp}, however, we find no evidence of a significant bias in the recovered signal. In the figure, we also demonstrate the MF's sensitivity to the signal template by considering different scale radii of $\theta_s \pm 0.2$ for both the FLlo and FLhi samples, showing that larger (smaller) $\theta_s$ will underestimate (overestimate) the true signals by around the $1\sigma$ error on the mean in the largest disk temperature bin, $\sim \pm7\mu K$.  This indicates the necessity of calibrating $\theta_s$ to extract an unbiased signal.

\begin{figure*}
\includegraphics[width = \linewidth]{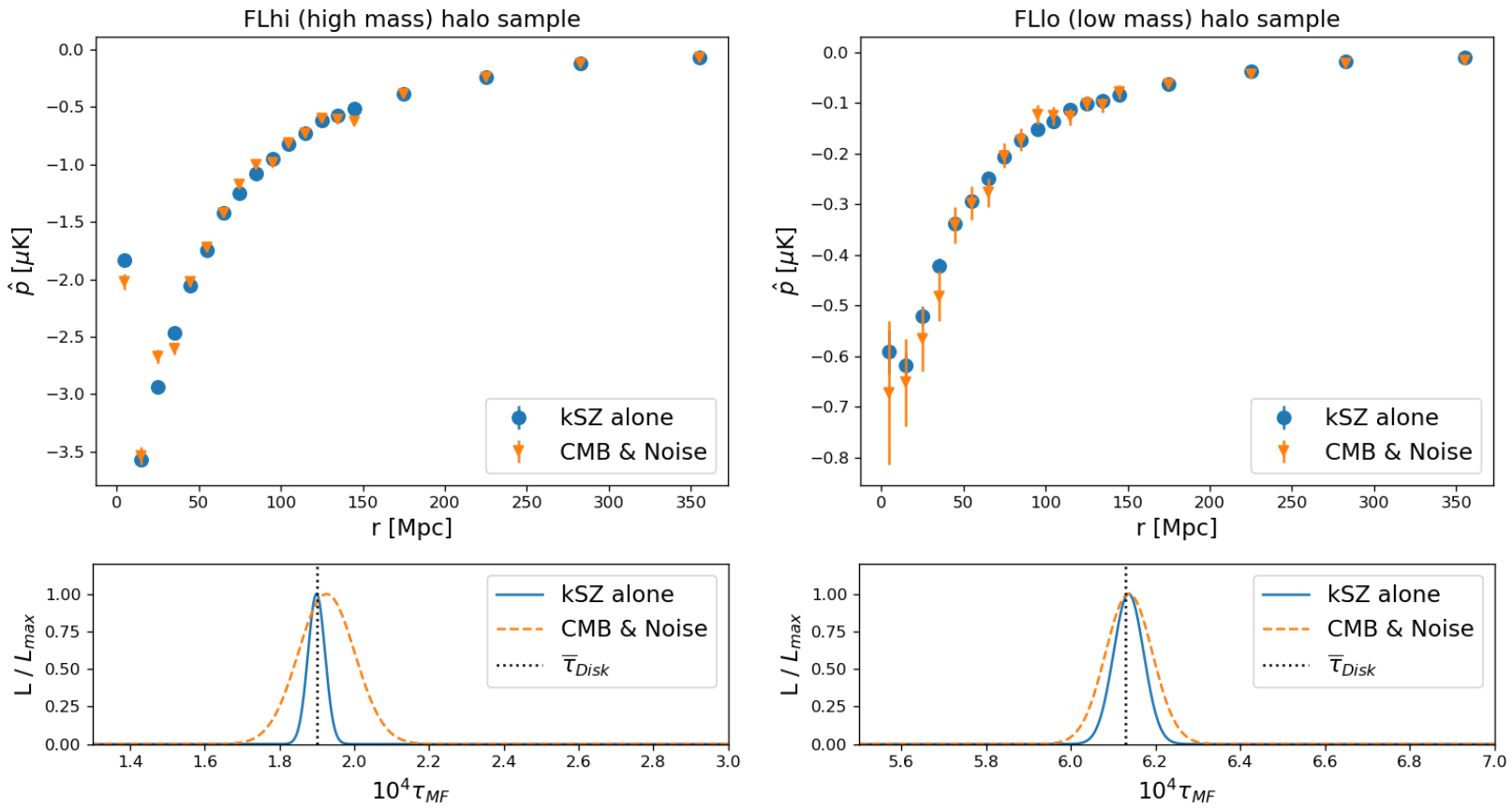}
\caption{The mean and 1-$\sigma$ error on the mean of the matched filter (MF) pairwise momentum estimator derived from the FL3 kSZ maps and 10 independent CMB+instrument noise realizations [orange] are shown with the pairwise statistic for the kSZ alone obtained from MF [blue] for the FLlo [left] and FLhi [right] halo samples with the ACT beam. [Lower] The normalized likelihoods for MF-obtained $ \tau$ estimate from kSZ alone [blue] and the mean from the 10 realizations with CMB and instrument noise included [orange].}
\label{fig:mf_result}
\end{figure*}

In summary, the application of the MF approach with these scale radii allows us to recover the disk kSZ signal for both the high and low mass cluster samples.

We use the calibrated MF template to analyze maps which include the 10 independent realizations of the CMB and instrument noise in combination with the FL3 kSZ map and convolved with the ACT beam (we also analyzed the Gaussian beam and found very similar results). As with AP analysis, large-scale signals ($\ell < 360$) are filtered out from the map before analyzing with MF, and the noise power spectrum P(k) in (\ref{eq:mf}) is calculated from the filtered map. 
The matched filtered temperatures, $T_{MF}$, include systematic residuals from the CMB and instrument noise. As discussed in section \ref{sec:AP_CMB_Noise}, however, we find that the pairwise momentum subtraction (\ref{eq:pairwise}) is efficient at removing systematic  
residuals not correlated with the pairwise in-fall velocities as originally constructed. 
 
In Fig.~\ref{fig:mf_result}, we show that the MF pairwise momentum estimators from the maps including CMB and noise recover an unbiased estimate of the kSZ signal. The effective $\tau_{MF}$ values recovered from the pairwise momentum curves show that the matched filter can effectively remove the primary CMB and instrument noise and recover an unbiased estimate of the optical depth for the kSZ signal.

We find that for the FLlo sample with ACT beam $10^4\overline{\tau}_{MF}=1.89\pm0.02 $ for the kSZ alone and $10^4\overline{\tau}_{MF}=1.93\pm0.06$ when primary CMB and instrument noise are added, with uncertainties three times larger. For the FLhi sample, the results are $\overline{\tau}_{MF}=6.14\pm0.03$ for the kSZ alone and $10^4\overline{\tau}_{MF}=6.14\pm0.05$ when primary CMB and instrument noise are added, with uncertainties two times larger. These results are consistent with the disk temperatures, $10^4\overline{\tau}_{disk} = 1.89\pm0.02$ and $10^4\overline{\tau}_{disk} = 6.13\pm0.03$ for the FLlo and FLhi samples respectively derived from the pure kSZ maps. The errors from this MF method are comparable to those from the AP approach where a fractional error of 4\% and 1\% is found for the FLlo and FLhi respectively. We find that FLhi has a comparably smaller uncertainty relative to FLlo due to its greater signal amplitude. 

We find the S/N = 10.3 at 2.1$^\prime$ for the MF results, averaged over the 10 noise realizations. This is comparable with that obtained from AP in Section~\ref{sec:ap_result_1}, consistent with \citep{Flender:2015btu}, in which one noise realization was used.

We note that in translating this approach to analyzing observational data,  the relative S/N obtained from  MF and AP approaches can be frequency dependent \citep{Planck:2011ai, Erler:2017dok}. At the frequencies of ACT observations, which provide the context for this work, the S/N from aperture photometry and matched filter approaches have been found to be consistent at the 1$\sigma$ level \citep{Erler:2017dok}.
 
We can also conduct a cross-check comparison with the optical depths obtained by fitting the individual halo temperature decrements to the halo LOS peculiar velocities. We find $10^4\overline{\tau}_{MF}=1.96\pm0.09$ and $10^4\overline{\tau}_{MF}=6.13\pm0.09$ for the FLlo and FLhi samples respectively when primary CMB and instrument noise are added, consistent with the pairwise-derived optical depth.

\subsection{Application to ACT data}
\label{sec:ACT_results}

\begin{figure}
\includegraphics[width = \columnwidth]{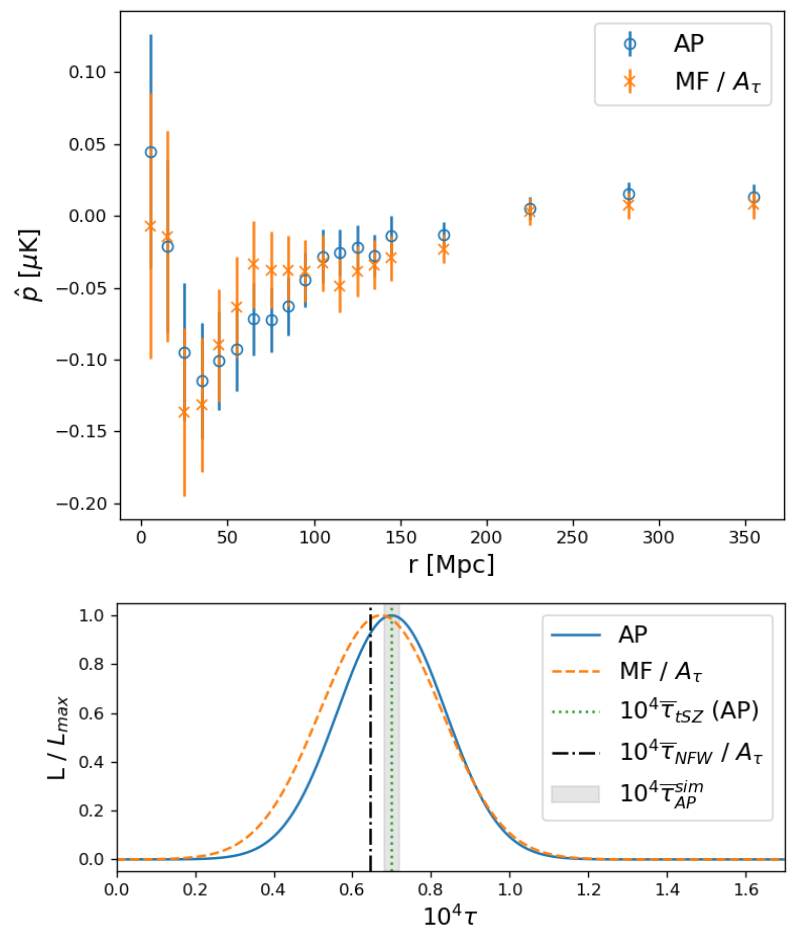}
\caption{[Upper] The mean and 1-$\sigma$ error of the AP [blue] $\&$ MF [orange] pairwise momentum estimator derived from the ACT DR5 observations\citep{Naess:2020wgi} for the L61 halo samples\citep{Calafut:2021wkx}. [Lower] The normalized likelihoods for AP [blue] $\&$ MF [orange]-obtained $\ \tau$ estimates along with that obtained from the aperture photometry analysis of the FL3 and FLlo simulations ($10^4\overline{\tau}_{AP}^{sims} =0.70\pm0.02$) and tSZ derived estimates using the $\overline{y}_{AP}-\overline{\tau}_{AP}$ relation in (\ref{eq:ytau}). The FL3 FLlo AP attenuation factor, $A_\tau=2.74$, has been applied to the MF to account for the difference between the AP amplitude, after annulus subtraction, and that of the disk itself (which the MF estimates).}
\label{fig:L61}
\end{figure}

We complete the analysis by implementing the calibrated aperture photometry and matched filter techniques, developed in sections \ref{sec:ap_result} and \ref{sec:mf_result}, to the ACT DR5 L61 sample that was analyzed in \citep{Calafut:2021wkx, Vavagiakis:2021ilq} to obtain kSZ pairwise momentum predictions and comparing them with the tSZ derived y-parameter. 

We use the FL3 simulation as the basis to determine the aperture photometry multiplicative factor, $A_\tau$, and to calibrate the matched filter. As discussed, the Flender FL3 simulation has a comparable resolution to the ACT maps (0.43$^\prime$ and 0.5$^\prime$ respectively) and includes a detailed baryonic model, including the effects of AGN feedback, star formation, and radially dependent non-thermal pressure support. The FLlo sample was selected to have comparable maximum and mean masses and mean redshift, to the L61 sample. 

We infer optical depth estimates from the pairwise momenta by comparing them to a theoretical pairwise velocity estimate obtained assuming a $Planck$ cosmology for a flat universe \citep{Planck:2015fie} as in \citep{Calafut:2021wkx}. This uses a halo mass function in \citep{Bhattacharya:2010wy} and a modified version of the CAMB code \citep{Lewis:1999bs} to calculate the mass-averaged pairwise velocity, as described in Mueller et al. \citep{Mueller:2014nsa, Mueller:2014dba}.

In Fig.~\ref{fig:L61}, we present the pairwise momentum kSZ signals and $\tau$ estimates recovered with the AP and MF techniques. From the AP approach we find $10^4\overline{\tau}_{AP} =0.69\pm0.11$ while for the MF method $10^4\overline{\tau}_{MF} = 1.84\pm 0.37$. While the MF approach estimates the optical depth with the 2.1$^\prime$ disk, the AP pairwise momentum includes the signal reduction, relative to the estimated disk signal, from the annulus subtraction. We use the attenuation factor calibrated from FL3-FLlo sample, $A_\tau=2.74\pm0.03$ to address this in the analysis here. As aperture photometry is the most commonly used method and in putting direct connection to the original ACT DR5 analysis using AP in \citep{Calafut:2021wkx, Vavagiakis:2021ilq}, we leave the aperture photometry results unchanged and employ the attenuation factor to the disk measurements to show the connection and consistency.

Using this, we find that the estimated optical depths from the aperture photometry and the calibrated matched filter technique, $10^4\overline{\tau}_{AP}^{ACT}=0.69\pm0.11$ and $10^4\overline{\tau}_{MF}^{ACT}/A_\tau  =0.67\pm 0.14$, respectively, are in extremely good agreement with one another. These relate to estimates of the total optical depth within $2.1^\prime$, $\overline{\tau}_{disk}$, of $10^4 A_{\tau} \times \overline{\tau}_{AP} = 1.89 \pm 0.31$ and $10^4\overline{\tau}_{MF} = 1.84\pm 0.37$ for AP and MF respectively.

We find the tSZ results are also consistent with the tSZ-derived optical depth estimates if we use the simulation calibrated $\overline{y}_{AP}-\overline{\tau}_{AP}$ relation presented in 
(\ref{eq:ytau}). Using this, the tSZ L61 result of $10^7\bar{y}_{AP}^{ACT}=0.79 \pm 0.11$, reported in \citep{Vavagiakis:2021ilq}, implies a tSZ AP-derived optical depth of $10^4\overline{\tau}_{tSZ}^{ACT}=0.70\pm0.06$, including the errors on $\alpha, \overline{y}_0$ and $\overline{y}_{AP}$.
 
The values obtained with the ACT data are also consistent with those obtained from 10 CMB+noise realizations with the FLlo sample and FL3 kSZ model, for which $10^4\overline\tau_{AP}^{sim} =0.70\pm0.02$ and calibrated matched filter technique,  $10^4\overline\tau_{MF}^{sim} =1.93 \pm 0.06$ ($10^4\overline\tau_{MF}^{sims}/A_\tau =0.70\pm 0.02$). Within current error margins, this is also consistent with the NFW-based optical depth estimate within 2.1$^\prime$ from \citep{Battaglia:2016xbi} used for the comparison in \citep{Vavagiakis:2021ilq},  $10^4\overline{\tau}_{NFW}=1.77 $ (and $10^4\overline{\tau}_{NFW}/A_\tau =0.65$). 

The uncertainties on the ACT $\tau$ estimates are larger than those from the simulation. This is clearly to be expected as there are additional uncertainties in the observational error budget that are not present in the simulation-derived results. As one example, the cluster mass estimates for the L61 sample are inferred from a multi-step process: the observed central galaxy luminosity is used to estimate the total cluster luminosity which, in turn, is used to infer the cluster halo mass \citep{Vavagiakis:2021ilq}. Each stage of this inherently adds uncertainties relative to the simulations, for which the halo masses are known. A second origin of uncertainty is the mis-centering of the target galaxy\citep{Flender:2015btu, Calafut:2017mzp} which would create an inherent offset for the centering of both the aperture photometry and matched filter signal extraction approaches.  

\section{Conclusion}
\label{sec:conc}

In this work, we have determined how well the pairwise kSZ momentum can be used to recover the kSZ signal in galaxy clusters, and how it can then be used to accurately infer an associated optical depth. Two different and complementary filtering approaches are used: aperture photometry and a calibrated matched filter. The sensitivity of the pipeline to the cluster sample masses and the effects of baryonic clustering are studied. We consider two halo samples and three kSZ models available from the simulations in  Flender et al. \cite{Flender:2015btu}. One of the halo samples (FLlo) includes halos down to a mass of $10^{13} M_{\odot}$ and is created so as to have a mean halo masses and a mean redshift comparable to that of the SDSS DR15 sample used for the ACT DR5 L61 sample analyzed in \cite{Vavagiakis:2021ilq, Calafut:2021wkx}. The other (FLhi) focuses on high mass clusters, $M>10^{14} M_{\odot}$.  The three simulated kSZ maps, referred to here as FL1-FL3, are created under different modeling assumptions for the gas prescription. Model FL1 assumes baryon traces the dark matter while Models FL2 and  FL3 use a gas prescription from Shaw et al. (2010)\cite{Shaw:2010mn} with FL3 also including diffuse gas. We perform the signal extraction after filtering out large-scale CMB modes ($\ell<360$) in the maps, and smoothing them with the inclusion of a Gaussian beam (FWHM = 1.4$^\prime$) or the ACT Beam from ACT DR5 \cite{Naess:2020wgi}, and noise realizations modeling the ACT instrument noise. We obtain an inferred best-fit mass-averaged optical depth, $ \tau$, by comparing the extracted pairwise momentum signal to the simulated pairwise peculiar velocities, with a covariance obtained from a bootstrap analysis. 

We first consider the implications for kSZ measurements of extraneous signals from proximate correlated halos within the kSZ correlation length, the two-halo term, from co-aligned groups and clusters, and from diffuse gas along the line of sight centered on target halos within the catalog. We find the contribution of the two-halo term is only not negligible for scales above $2R_{200}$, and thus does not bias our measurements. On average, we find, in the FLlo catalog, that there are around two halos along the line of sight of each target halo within the 2.1$^\prime$ aperture. By considering the effect of adding further random co-aligned halos, we show that pairwise subtraction is effective in removing such systematic signal contamination and recovers an unbiased measurement of the pairwise signal. The diffuse gas component, on the contrary,  systematically enhances the pairwise signal, through its correlation with the primary halo, and does need to be accounted for.

Aperture photometry excises some of the cluster kSZ signals for the aperture sizes we use (motivated by signal-to-noise considerations). Using the simulations, we quantify the level of kSZ signal subtraction created when the annulus is subtracted from the disk signal. The accurate determination of the attenuation depends on the gas model, aperture size, map resolution, and beam convolution. We show that an attenuation factor calibrated off the pure-kSZ pairwise correlations can be used to relate the AP measurement from maps which include instrument noise and primary CMB to an unbiased estimate of the disk pairwise kSZ signal. 

We then compute the average tSZ Compton-y parameters and measure the corresponding average optical depth from kSZ maps using aperture photometry. These estimates are then used to model a $\overline{y}_{AP}-\overline{\tau}_{AP}$ scaling relation using AP-derived observables across the mass range sampled by the FLlo and FLhi catalogs. We demonstrate that using the tSZ measurements with AP and the $\overline{y}_{AP}-\overline{\tau}_{AP}$ relation for optical depth allows the inference of an unbiased estimate of the optical depth compared with the pairwise kSZ measurements using AP.

While aperture photometry is the principal approach used to date to conduct pairwise analyses, the need to accurately characterize the signal attenuation from annulus subtraction motivates considering other alternative techniques to cross-check. Here we propose a calibrated matched filter approach and show that it can provide a complementary method to aperture photometry to measure the kSZ signal. The matched filter is typically used to identify peaks embedded in a noise background without conserving amplitude. Here, we have used the projected NFW profile as the template signal and calibrated the scale angle, $\theta_s$, to ensure that the temperature obtained with the matched filtered matches that of the kSZ signal within the same 2.1$^\prime$ disk (i.e. $T_{MF}(\theta_s) =  T_{disk}$).  We demonstrate that this matched filtered approach works effectively to recover an unbiased estimate of the pure-kSZ pairwise signal and optical depth from maps including primary CMB and instrument noise. 

In \cite{Vavagiakis:2021ilq, Calafut:2021wkx}, the mass-averaged optical depth derived from the aperture photometry pairwise kSZ measurements of the ACT DR5 f150 data was noted to be smaller than the theoretical NFW-based prediction for the disk. We have shown that the difference can be resolved by factoring in the signal subtracted by the annulus for the 2.1$^\prime$ aperture, and have carefully calibrated this using simulations. We also show that the AP and MF analyses provide complementary estimates of the cluster optical depth that are both consistent with each other and consistent with that predicted from the Flender simulations for the FLlo halo sample, selected to cover a similar redshift and mass range and the NFW-based estimate. We conclude and propose the feasibility of using the AP attenuation factor(i.e. $A_\tau$) and the MF signal template(i.e. $\theta_s$) carefully calibrated from the simulations to analyze the real data.

We have demonstrated that the pairwise kSZ estimator can be used to accurately recover cluster properties such as the optical depth using two complementary filtering techniques as originally constructed. This includes an additional step, factoring in the annulus signal subtraction, for the widely used aperture photometry approach. For the matched filter approach, we show that calibration of the signal template profile can provide an effective alternative method.

The modeling of baryonic physics has an impact overall on the signal extraction method used in this work. Since the kSZ temperature profile varies under different baryonic physics assumptions, the AP attenuation factor, the MF signal template, and the $\overline{y}_{AP}-\overline{\tau}_{AP}$ relation discussed in this work are all calibrated on cluster modeling under the specific assumption of baryonic physics. We have considered baryonic modeling based on the Shaw model\citep{Shaw:2010mn}, in which the gas initially follows the dark matter density distribution(NFW profile\cite{Bartelmann:1996hq}) and then follows the hydrostatic equilibrium model. This modeling includes the effects of star formation, non-thermal pressure support, and AGN feedback. As improvements in cluster hydrodynamical gas modeling continue to develop (e.g. the characterization of gas cooling \citep{Flender:2016cjy}), there will be further opportunities to refine these techniques to model the gas physics and extract signatures of the underlying cosmological model.

This approach paves the way for future SZ measurements from CMB observations. For different cluster samples, we can use simulations with similar redshift, mass ranges, and map resolutions to obtain estimates of the AP attenuation factor and MF signal template profile. The connection between the simulations and observations can continue to be refined in future work through, for example, improvements in the baryonic modeling in the simulations and characterization of the potential impact of miscentering and errors in cluster mass estimation, currently inferred from the proxy galaxy luminosity.

The findings show the potential for the kSZ pairwise statistic to be a sensitive measure of cosmological clustering, and tests of gravity and the dark sector, as we prepare for upcoming analyses with the complete ACT dataset, the Simons Observatory, the CCAT Observatory, CMB-S4, and spectroscopic datasets from DESI, Euclid, and the Roman Space Telescope.

\section*{Acknowledgements}

We wish to thank Neelima Sehgal and Colin Hill for their helpful discussions in the course of this work. We also thank Bruce Partridge for his thoughtful and valuable comments on this paper as it was being drafted. The work of YG and RB is supported by NSF grant AST-2206088, NASA ATP grant 80NSSC18K0695, and NASA ROSES grant 12-EUCLID12-0004. EMV acknowledges support from NSF award AST-2202237. NB acknowledges the support from NASA ATP grant 80NSSC18K0695 and acknowledges additional support from NSF grant AST-1910021 and NASA grants 80NSSC22K0410 and 80NSSC22K0721. MDN acknowledges support from NSF grant AST-2117631.

\bibliographystyle{apsrev}
\bibliography{draft_prd_response}
\end{document}